\documentclass[%
reprint,
superscriptaddress,
twocolumn
preprintnumbers,
 amsmath,amssymb,
 aps,
]{revtex4-2}

\usepackage{graphicx}
\usepackage{dcolumn}
\usepackage{bm}
\usepackage{xcolor}

\begin{document}

\preprint{CERN-EP-2022-058}

\title{The $\pi^+\pi^-$ Coulomb interaction study and its use in the data processing} 

\author{B.Adeva}
\affiliation{Santiago de Compostela University, Spain}
\author{L.Afanasyev} 
\affiliation{JINR Dubna, Russia}
\author{A.Anania}
\affiliation{Messina University, Messina, Italy}
\author{S.Aogaki}
\affiliation{IFIN-HH, National Institute for Physics and Nuclear Engineering, Bucharest, Romania}
\author{A.Benelli} 
\affiliation{Czech Technical University in Prague, Czech Republic}
\author{V.Brekhovskikh}
\affiliation{IHEP Protvino, Russia}
\author{T.Cechak}
\affiliation{Czech Technical University in Prague, Czech Republic}
\author{M.Chiba} 
\affiliation{Tokyo Metropolitan University, Japan}
\author{P.Chliapnikov}
\affiliation{IHEP Protvino, Russia}
\author{D.Drijard}
\affiliation{Czech Technical University in Prague, Czech Republic}
\affiliation{CERN, Geneva, Switzerland}
\author{A.Dudarev}
\affiliation{JINR Dubna, Russia}
\author{D.Dumitriu} 
\affiliation{IFIN-HH, National Institute for Physics and Nuclear Engineering, Bucharest, Romania}
\author{P.Federicova}
\affiliation{Czech Technical University in Prague, Czech Republic}
\author{A.Gorin}
\affiliation{IHEP Protvino, Russia}
\author{K.Gritsay}
\affiliation{JINR Dubna, Russia}
\author{C.Guaraldo}
\affiliation{INFN, Laboratori Nazionali di Frascati, Frascati, Italy}
\author{M.Gugiu}
\affiliation{IFIN-HH, National Institute for Physics and Nuclear Engineering, Bucharest, Romania}
\author{M.Hansroul}
\affiliation{CERN, Geneva, Switzerland}
\author{Z.Hons}
\affiliation{Nuclear Physics Institute ASCR, Rez, Czech Republic}
\author{S.Horikawa}
\affiliation{Zurich University, Switzerland}
\author{Y.Iwashita}
\affiliation{Kyoto University, Kyoto, Japan}
\author{J.Kluson} 
\affiliation{Czech Technical University in Prague, Czech Republic}
\author{M.Kobayashi}
\affiliation{KEK, Tsukuba, Japan}
\author{L.Kruglova} 
\affiliation{JINR Dubna, Russia}
\author{A.Kulikov} 
\affiliation{JINR Dubna, Russia}
\author{E.Kulish}
\affiliation{JINR Dubna, Russia}
\author{A.Lamberto} 
\affiliation{Messina University, Messina, Italy}
\author{A.Lanaro}
\affiliation{University of Wisconsin, Madison, USA}
\author{R.Lednicky} 
\affiliation{Institute of Physics ASCR, Prague, Czech Republic }
\author{C.Mari\~nas}
\affiliation{Santiago de Compostela University, Spain}
\author{J.Martincik}
\affiliation{Czech Technical University in Prague, Czech Republic}
\author{L.Nemenov}
\thanks{Corresponding author}
\email{nemenov@cern.ch}
\affiliation{JINR Dubna, Russia}
\author{M.Nikitin}
\affiliation{JINR Dubna, Russia}
\author{K.Okada} 
\affiliation{Kyoto Sangyo University, Kyoto, Japan}
\author{V.Olchevskii}
\affiliation{JINR Dubna, Russia}
\author{M.Pentia} 
\affiliation{IFIN-HH, National Institute for Physics and Nuclear Engineering, Bucharest, Romania}
\author{A.Penzo}
\affiliation{INFN, Sezione di Trieste, Trieste, Italy}
\author{M.Plo}
\affiliation{Santiago de Compostela University, Spain}
\author{P.Prusa}  
\affiliation{Czech Technical University in Prague, Czech Republic}
\author{G.Rappazzo} 
\affiliation{Messina University, Messina, Italy}
\author{A.Romero Vidal}
\affiliation{Santiago de Compostela University, Spain}
\author{A.Ryazantsev}
\affiliation{IHEP Protvino, Russia}
\author{V.Rykalin}
\affiliation{IHEP Protvino, Russia}
\author{J.Saborido} 
\affiliation{Santiago de Compostela University, Spain}
\author{J.Schacher}
\affiliation{Albert Einstein Center for Fundamental Physics, Laboratory of High Energy Physics, Bern, Switzerland}
\author{A.Sidorov}
\affiliation{IHEP Protvino, Russia}
\author{J.Smolik}
\affiliation{Czech Technical University in Prague, Czech Republic}
\author{F.Takeutchi} 
\affiliation{Kyoto Sangyo University, Kyoto, Japan}
\author{T.Trojek}
\affiliation{Czech Technical University in Prague, Czech Republic}
\author{S.Trusov} 
\affiliation{Skobeltsyn Institute for Nuclear Physics of Moscow State University, Moscow, Russia}
\author{T.Urban}
\affiliation{Czech Technical University in Prague, Czech Republic}
\author{T.Vrba}
\affiliation{Czech Technical University in Prague, Czech Republic}
\author{V.Yazkov} 
\thanks{deceased}
\affiliation{Skobeltsyn Institute for Nuclear Physics of Moscow State University, Moscow, Russia}
\author{Y.Yoshimura} 
\affiliation{KEK, Tsukuba, Japan}
\author{P.Zrelov} 
\affiliation{JINR Dubna, Russia}




\collaboration{DIRAC Collaboration}

\date{\today}


\begin{abstract}
In this work the Coulomb effects (Coulomb correlations) 
in $\pi^+\pi^-$ pairs produced in p + Ni collisions at 
24 GeV/$c$, are studied 
using experimental $\pi^+\pi^-$ pair distributions in $Q$, the relative 
momentum in the pair center of mass system (c.m.s), and its projections 
$Q_L$ (longitudinal component) and $Q_t$ (transverse component) 
relative to the pair direction in the laboratory system (l.s.). 
The major part of the pion pairs ({\sl Coulomb pairs}) is produced in the decay of $\rho, \omega$ and $\Delta$-resonances and other short-lived sources. 
In these pairs, the significant Coulomb interaction occurs at small $Q$, 
dominating the $\pi^+\pi^-$ interaction in the final state.

The minor part of the pairs ({\sl non-Coulomb pairs}) is produced 
if one or both pions arose from long-lived sources like $\eta, \eta'$ or from different interactions.
In this case, the final state interaction is practically absent. 

The $Q$, $Q_L$, and $Q_t$ distributions of the {\sl Coulomb pairs} in the c.m.s. have been simulated assuming they are described by the phase space modified by the known point-like Coulomb correlation function $A_C(Q)$, corrected for small effects due to the nonpoint-like pair production and the strong two-pion interaction. 
The same distributions of {\sl non-Coulomb pairs} have been simulated according to the phase space, but without $A_C(Q)$.

In all $Q_t$ intervals, the experimental $Q_L$ spectrum shows a peak around $Q_L = 0$ caused by the Coulomb final state interaction. 
The full width at half maximum increases with $Q_t$ from 3~MeV/$c$ 
for $0<Q_t<0.25$~MeV/$c$ to 11~MeV/$c$ for $4.0<Q_t < 5.0$~MeV/$c$. 
The experimental $Q_L$ distributions have been fitted with two free parameters: the fraction of {\sl Coulomb pairs} and the normalization constant. The precision of the description of these distributions is better than $2\%$ in $Q_t$ intervals 2 -- 3, 3 -- 4, and 4 -- 5 MeV/$c$, and better than $0.5\%$ in the total $Q_t$ interval 0 -- 5 MeV/$c$.

It is shown that the number of {\sl Coulomb pairs} in all $Q_t$ intervals, 
including the small $Q_t$ (small opening angles $\theta$ in the l.s.)
is calculated with the theoretical precision better than 2\%. 

The comparison of the simulated and experimental numbers of {\sl Coulomb pairs} at small $Q_t$ allows us to check and correct the detection efficiency for the pairs with small $\theta$ (0.06 mrad and smaller).

It is shown that {\sl Coulomb pairs} can be used as a new physical tool 
to check and correct the quality of the simulated events. The special property of the {\sl Coulomb pairs} is the possibility of checking and correcting the detection efficiency, especially for the pairs with small opening angles. 
\vspace{-5mm}

\end{abstract}

    

\maketitle



\section{Introduction}

The Coulomb interaction effect was first observed and investigated in the hadron pair production in \cite{AFAN91}. The pairs were produced in the reaction
\vspace{-4mm}

\begin{equation}
\label{p-Ta-interac}
p+{\rm Ta} \longrightarrow \pi^+ \pi^- + X 
\end{equation}
at the proton momentum of 70~GeV/$c$. 
The generation of the pairs ({\sl {\sl Coulomb pairs}}) was described \cite{NEME85}
as the product of the pair production matrix element without Coulomb interaction in the final state and the Coulomb correlation function $A_C(Q)$ \cite{GAMO28,SOMM31,SAKH48,SAKH91}, where $Q$ is the relative momentum in the pair center of mass system (c.m.s). This approach was used by analogy with the theoretical description of the Coulomb final state interaction in $e^+e^-$ pair production in photon-nucleus interaction \cite{SAKH91}.
In both cases the pair production region $\sim 1/m$ ($m$ is the pion or electron mass) is  two orders of magnitude smaller than the distance $R\sim 1/\alpha m$ ($\alpha=1/137$ is the fine structure constant) over which the wave function of the relative motion of the particles changes. It allows one to use the wave function value at $r=0$. 

The $Q$ distribution \cite{AFAN91} of $\pi^+\pi^-$ {\sl pairs} produced in one p+Ta interaction ({\sl prompt pairs}) $F(Q)_{pr}$ was divided by the same distribution $F(Q)_{acc}$ of the {\sl accidental pairs} generated at two different production points, without interaction in the final state. 
This ratio $R(Q)_\mathrm{exp}=F(Q)_{pr}/F(Q)_{acc}$ is normalized to unity at the large $Q$ and describes, by definition, the $\pi^+\pi^-$ final state interaction dependence on $Q$ (Coulomb correlation function).

The theoretical ratio $R(Q)_\mathrm{calc}$ was evaluated with the phase space restrictions due to setup acceptance and Coulomb interaction in the final state.
It was shown that $R(Q)_\mathrm{calc}$ described well the ratio $R(Q)_\mathrm{exp}$ in the total ana\-lyzed $Q$ interval 0 -- 40~MeV/$c$. 
The Coulomb correlation function increases when $Q$ decreases. 
In (\ref{p-Ta-interac}) the $R(Q)_\mathrm{exp}$ value increased 6 times when $Q$ decreased from 40~MeV/$c$ to 0.5~MeV/$c$.
The function $R_{exp}$ dependence on $Q_L$ (longitudinal 
component) and $Q_t$ (transverse component) was also well described.

The $Q_L$ distribution of the prompt $\pi^+\pi^-$ pairs was analyzed using the following procedure. For an experimental accidental pion pair with $Q_L$, the weight $A_C(Q_L)$ was introduced to ”create” a {\sl Coulomb pair}. {\sl Coulomb pairs} are generated when $\pi^+$  and $\pi^-$  are produced from the decay of $\rho, \omega, \Delta$ and other short-lived sources. If one or both pions are produced from long-lived sources like $\eta, \eta'$ or $K^0$'s, then the distance between particles is larger and the Coulomb interaction in the final state is almost absent. These pairs were defined as ”non-Coulomb” pairs (”decay pairs” in \cite{AFAN91}) and their distribution in $Q_L$ was the same as the $Q_L$ spectrum of the accidental pairs. The experimental $Q_L$ distribution was described by the sum of the {\sl Coulomb} and {\sl non-Coulomb pairs}. The ratio between the {\sl Coulomb} and {\sl non-Coulomb pairs} was taken from the Lund model \cite{LUND83}. The experimental $Q_L$ spectrum was well fitted with one free parameter - the normalization constant. 

The Coulomb effects in the $\pi^+\pi^-$ and $p\pi^-$  pairs were observed and described in \cite{WIEN92}. 

\begin{figure}[h]
\centering
\includegraphics[width=80mm]{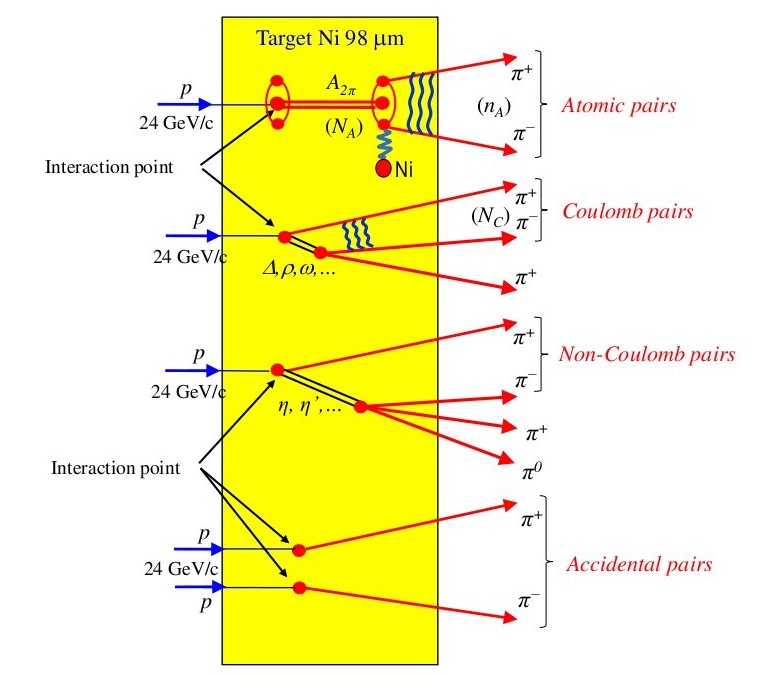}
\caption{
The atomic, Coulomb, non-Coulomb, and accidental pair production processes. 
The wave lines denote the Coulomb interaction.}
\label{fig:atom-production}
\end{figure}

The pairs with the Coulomb interaction in the final state create the main background for the observation and investigation of the $\pi^+\pi^-$ atoms \cite{NEME85}. Therefore, to observe $\pi^+\pi^-$  atoms, the {\sl Coulomb pair} distributions must be described accurately. A detailed description of the $\pi^+\pi^-$  pair spectrum was given in \cite{AFAN93}, where $\pi^+\pi^-$ atoms were observed for the first time. The $\pi^+\pi^-$ pairs in proton-nucleus interaction are produced on any target in the processes shown in Fig.\ref{fig:atom-production} (Ni target).
\vspace{-2mm}

The $\pi^+ \pi^-$ atoms produced in the p+Ta interaction are broken up (ionized) with a large probability while moving in the target, which results in generation of $\pi^+\pi^-$ pairs ({\sl atomic pairs}).
Since {\sl non-Coulomb} and {\sl accidental pairs} are uncorrelated and indistinguishable in the momentum space, we call here their sum  {\sl non-Coulomb pairs}. 
The relative momentum $Q$ of {\sl atomic pairs} is less than 3 MeV/$c$ for thin targets ($10^{-3} X_0$). Due to this specific kinematical feature, {\sl atomic pairs} are experi\-mentally observable. 
These pairs number $n_A$ we need to measure the $\pi^+ \pi^-$ atom lifetime and the $\pi^+ \pi^-$ scattering length in the $s$-state. 
To evaluate the $n_A$ number, the {\sl Coulomb} and {\sl non-Coulomb pairs} $(Q, Q_L)$ distributions are described in a wide interval of these parameters. Then the fitting distributions are subtracted from the total $(Q, Q_L)$ spectrum. The precision of the {\sl Coulomb pair} spectrum defines the accuracy of $n_A$ and the error of the $\pi^+ \pi^-$ scattering length.

The distributions in $Q_L$ and the $Q_t$ components $Q_x$ and $Q_y$ are Gaussian-like and have different standard deviations (s.d.) $\sigma_L, \sigma_X$ and $\sigma_Y$. 
Therefore, the prompt $\pi^+\pi^-$ distribution was analyzed using 
the parameter

\begin{equation}
\label{F-error}
F=\sqrt{
\left(\frac{Q_L}{\sigma_L}\right)^2
+\left(\frac{Q_X}{\sigma_X}\right)^2
+\left(\frac{Q_Y}{\sigma_Y}\right)^2
}
\end{equation}

The {\sl Coulomb} and {\sl non-Coulomb pair} distributions as the function of $F$ were obtained in the same way as the pair distributions in $Q_L$ described above. 
It was shown that the prompt pair spectrum in $F$ in the interval 0--40~MeV/$c$ is well described as the sum of the {\sl Coulomb} and {\sl non-Coulomb pairs} distribution with two free parameters: the normalization constant and the ratio {\sl Coulomb} to {\sl non-Coulomb pairs}.

A more precise description of the {\sl Coulomb} and {\sl non-Coulomb pairs} was done in the DIRAC experiment \cite{ADEV05} at CERN in the measurement of the $\pi^+\pi^-$ atom lifetime and the $\pi\pi$ scattering length.
In this experiment $\pi^+\pi^-$, $\pi^+K^-$, $\pi^-K^+$, $K^+K^-$ and 
$p\overline p$ hadron pairs were generated in the process 
\vspace{-4mm}

\begin{equation}
\label{p-Ni-interac}
p+{\rm Ni} \longrightarrow h^+ h^- + X
\end{equation}

with the proton momentum of 24~GeV/$c$.

The $Q$ distribution of {\sl Coulomb pair} was simulated
assuming they are described by the phase space modified by the Coulomb correlation function $A_C(Q)$. The same spectrum of the {\sl non-Coulomb pairs} was simulated (without $A_C(Q)$). The c.m.s. pion momenta were transformed to the l.s. using the experimental total momentum of the $\pi^+\pi^-$ pairs. 
The difference between the total momentum distribution of the {\sl Coulomb} and {\sl non-Coulomb pairs} was taken into account using FRITIOF-6 code 
\cite{FRIT87}.
This approach allowed a good description of the $Q$ and $Q_L$ distributions. In the second DIRAC experiment \cite{ADEV11} larger experimental data were analyzed. 

{\sl Coulomb pair} simulation in \cite{ADEV11} included the Coulomb and strong $\pi^+\pi^-$ interactions in the final state and the influence of the nonpoint-like pair production on the spectrum shape \cite{LEDN08}.
The sources of the nonpoint-like {\sl Coulomb pair} production
were investigated in \cite{CHLI09}.
A new procedure \cite{ZHAB07,ZHAB08} was used, which more accurately took into account the difference between the total momentum distributions of the {\sl Coulomb} and {\sl non-Coulomb pairs} in the l.s. 
This analysis enabled a good description of the {\sl Coulomb pair} distribution in $Q_L$ and $Q_t$ in the intervals 0--15~MeV/$c$ and 0--5~MeV/$c$ respectively.

The DIRAC setup was upgraded to identify and investigate  $\pi^+\pi^-$,
$\pi^+K^-$, $\pi^-K^+$, $K^+K^-$ and $p\overline p$ pairs
\cite{ADEV16a}.
In the dedicated experiment \cite{ADEV17}, distributions of {\sl Coulomb}, {\sl non-Coulomb} and {\sl atomic} $\pi^+K^-$ and $\pi^-K^+$ pairs were accurately described. An improved version of the simulation procedure and a more accurate setup geometry tuning were used \cite{BENE16}.
\\
This allowed us to observe for the first time the $\pi^+ K^-$ and $\pi^- K^+$ atoms, to measure their lifetime and to evaluate the $\pi K$ scattering length. In all those investigations the $\pi^+\pi^-$ pairs as the background processes were used to check the setup tuning \cite{BENE16}.
The same simulation procedure was used in the present work.

The Coulomb interaction in the final state was studied both theoretically and experimentally in \cite{AFAN91, NEME85, GAMO28, SOMM31, SAKH48, SAKH91, WIEN92, AFAN93, ADEV05, ADEV11, LEDN08, CHLI09, ZHAB07, ZHAB08, ADEV16a, ADEV17} before our reported investigations. In this Introduction, we present the theoretical description of the Coulomb interaction and its use in analyzing the experimental results in the works cited above. 

The present work (see also \cite{BENE23}) deals with the principal new investigation of 
the $\pi^+\pi^-$ pairs detected by the upgraded DIRAC setup \cite{ADEV16a} with new detectors for suppression of $K$ mesons, protons, and antiprotons. 
It allowed 
one to decrease the admixture of $K^+K^-$ and $p\overline p$ pairs in the pion pair data.
It was shown that selection of two-pion {\sl Coulomb pairs} in sufficiently narrow $Q_t$ intervals makes it possible to control the width of the Coulomb peak of the $Q_L$ distribution around $Q_L=0$ and describe it with a precision better than $2\%$. This enables using the {\sl Coulomb pairs} as a new physical tool to check and correct the detector resolution and efficiency, especially for the pairs with opening angle $\theta$ in the l.s. down to about 0.02 mrad.

\section{Setup and experimental conditions} 
\label{sec:setup}
\begin{figure}[ht]
\begin{center}
\includegraphics[width=0.95\columnwidth]{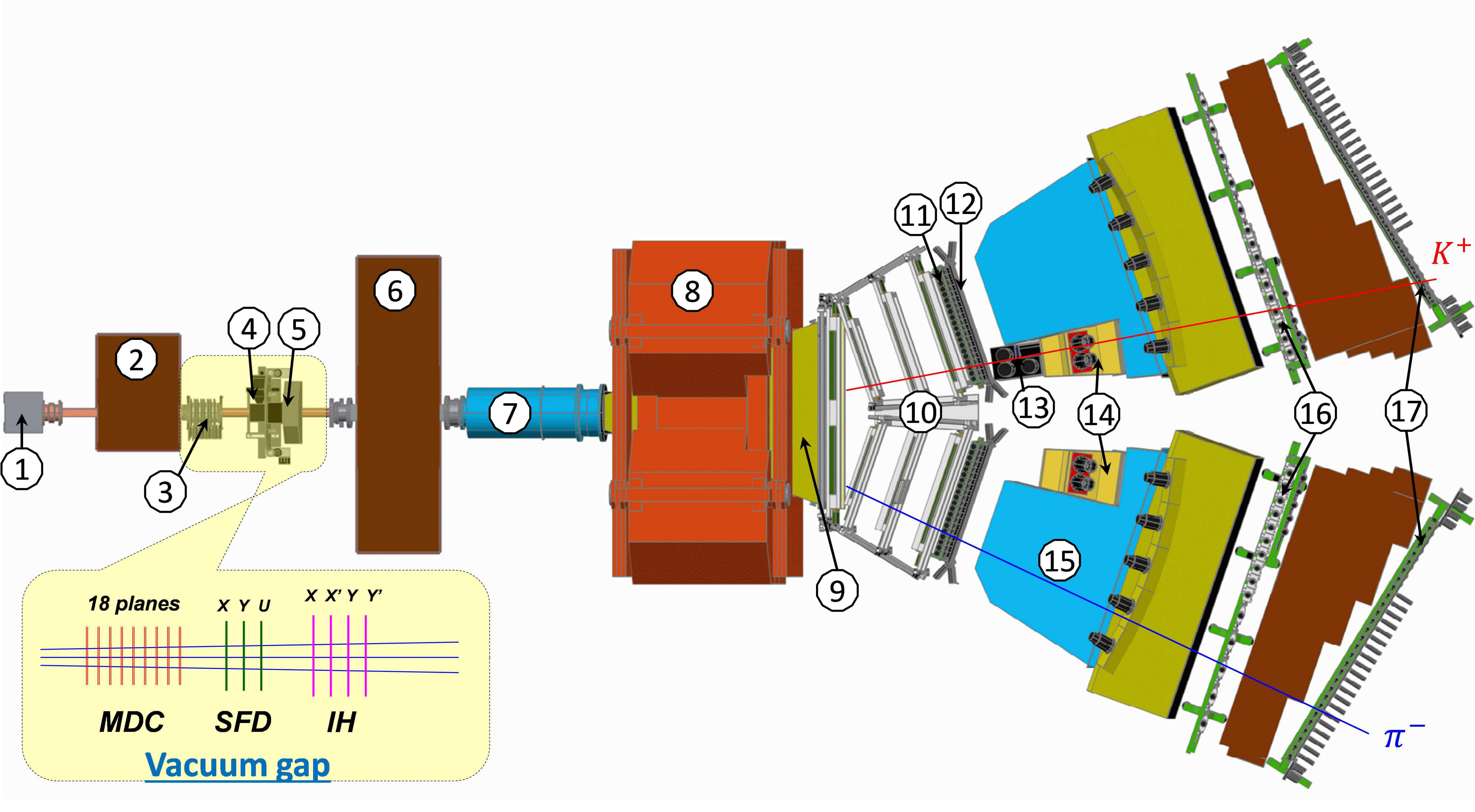}
\caption{ General view of the DIRAC setup 
(1 -- target station;
2 -- first shielding;
3 -- micro drift chambers (MDC);
4 -- scintillating fiber detector (SFD); 
5 -- ionization hodoscope (IH); 
6 -- second shielding; 
7 -- vacuum tube; 
8 -- spectrometer magnet; 
9 -- vacuum chamber; 
10 -- drift chambers (DC); 
11 -- vertical hodoscope (VH); 
12 -- horizontal hodoscope (HH); 
13 -- aerogel Cherenkov (ChA); 
14 -- heavy gas Cherenkov (ChF); 
15 -- nitrogen Cherenkov (ChN); 
16 -- preshower (PSh); 
17 -- muon detector (Mu).}
\label{fig:det}
\end{center}
\end{figure}

The aim of the magnetic two-arm vacuum spectrometer 
\cite{ADEV16a,GORC05a, GORC05b, PENT15} 
(Fig.~\ref{fig:det}) is to detect and identify  
$K^+ K^-$ , $\pi^+ \pi^-$ , $\pi^- K^+$, and $\pi^+ K^-$ 
pairs with small $Q$. The structure of 
$K^+ K^-$ and $\pi^+ \pi^-$  pairs downstream the magnet is 
approximately symmetric. The 24~GeV/$c$ primary proton beam, 
extracted from the CERN PS, hit a Ni target of the
$(108\pm 1)$\textmu{}m thickness or $7.4\cdot10^{-3} X_0$. 

The axis of the secondary channel is inclined 
relative to the proton beam by $5.7^\circ$ upward. The solid 
angle of the channel is $\Omega = 1.2 \cdot 10^{-3}$ sr. 
Secondary particles propagate mainly in vacuum to the Al foil 
$(7.6 \cdot 10^{-3} X_{0})$ at the exit of the vacuum chamber, 
which is installed between the poles of the dipole magnet 
($B_{max}$ = 1.65~T and $BL$ = 2.2~Tm).
In the vacuum channel gap, 18 planes of the Micro Drift Chambers (MDC) 
and ($X$, $Y$, $U$) planes of the Scintillation Fiber Detector (SFD) 
were installed to measure both the particle coordinates  
($\sigma_{SFDx} = \sigma_{SFDy} = 60$~\textmu{}m, 
$\sigma_{SFDu} = 120$~\textmu{}m) and the particle time 
($\sigma_{tSFDx} = 380$~ps, $\sigma_{tSFDy} = \sigma_{tSFDu} = 520$~ps).  
The total matter radiation thickness between the target and the vacuum chamber 
amounts to $7.7 \cdot 10^{-2} X_{0}$. 

Each spectrometer arm is equipped with the following subdetectors 
\cite{ADEV16a}: drift chambers (DC) to measure particle coordinates with 
approximately  $85~\mu m$ precision and to evaluate the particle path length; a vertical hodoscope (VH) to determine particle times with $110~$ps accuracy for identification of equal mass pairs via the time of flight (TOF) between the SFDx plane and the VH; a horizontal hodoscope (HH) to select particles with a vertical distance less than 75~mm ($Q_{Y}$ less than 15~MeV/$c$) in the two arms; an aerogel Cherenkov counter (ChA) to distinguish kaons from protons; a heavy gas ($\text{C}_{4} \text{F}_{10}$) Cherenkov counter (ChF) to distinguish pions from kaons and protons; a nitrogen Cherenkov (ChN) and preshower (PSh) detectors to identify and reject $e^+e^-$ pairs; an iron absorber and a two-layer scintillation counter (Mu) to identify muons. 
In the negative arm, no aerogel Cherenkov counter was installed, 
because the number of antiprotons is small compared to $K^-$. 

Pairs of oppositely charged particles, produced in one p+Ni interaction 
(prompt pairs) and accidentals produced in two different 
interactions in the time interval $\pm 20~$ns are selected by 
requiring a 2-arm coincidence (ChN in anticoincidence) with 
the coplanarity restriction (HH) in the first-level trigger. 
The second-level trigger selects events with at least one track 
in each arm by exploiting the DC-wire information (track finder). 
Particle pairs $\pi^{-} p$ ($\pi^{+} \bar{p}$) from 
$\Lambda$ ($\bar{\Lambda}$) decay were used for spectrometer calibration, and $\text{e}^+\text{e}^-$ pairs were employed for general detector calibration. 


\section{Data processing} 
\label{sec:dp}

The collected events were analyzed with the DIRAC reconstruction program 
ARIANE \cite{Ariane}. 

\subsection{Tracking} 
\label{ssec:track}

Only events with one or two particle tracks in the DC   
of each arm are processed. The event reconstruction is 
performed as follows \cite{ADEV17}:
\begin{itemize}
\item One or two hadron tracks are identified in the DC  
of each arm with hits in VH, HH, and PSh slabs and 
no signal in ChN and Mu.
\item Track segments reconstructed in the DC are extrapolated 
backward to the beam position in the target, 
using the transfer function of the dipole magnet and 
the program ARIANE. This procedure 
provides approximate particle momenta and  
the corresponding points of intersection in the MDC, SFD, and IH.
\item Hits are searched for around the expected SFD coordinates 
in the region $\pm 1$~cm corresponding to (3--5)~$\sigma_\text{pos}$ 
defined by the position accuracy with allowance for 
the particle momenta. 
To identify the event when two particles cross the same SFD co\-lumn, 
 the double ionization in the corresponding IH slab was requested. 
\end{itemize}
The momentum of the positively or negatively charged particle 
is refined to match the $X$-coordinates of the DC tracks 
as well as the SFD hits in the $X$- or $U$-plane, 
depending on the presence of hits. To find 
the best 2-track combination, the two tracks should not use 
a common SFD hit in the case of more than one hit in the proper region. 
In the final analysis, the combination with the best $\chi^2$ 
in the other SFD planes is kept. 

\subsection{Setup tuning with $\Lambda$ and $\bar{\Lambda}$ particles} 
\label{ssec:geom}

To check the general geometry of the DIRAC experiment, 
 $\Lambda$ and $\bar{\Lambda}$ particles decaying into 
${\rm p}\pi^-$ and $\pi^+\bar{\rm p}$ in our setup were 
used \cite{BENE16}. After the setup tuning
the weighted average of the experimental $\Lambda$ mass overall runs, 
$M^\mathrm{DIRAC}_{\Lambda} = (1.115680 \pm 2.9 \cdot 10^{-6})$~GeV/$c^2$, 
agrees very well with the PDG value,  
$M^\mathrm{PDG}_{\Lambda} = (1.115683 \pm 6 \cdot 10^{-6})$~GeV/$c^2$. 
The weighted average of the experimental ${\bar{\Lambda}}$ mass is 
$M^\mathrm{DIRAC}_{\bar{\Lambda}} = (1.11566 \pm 1 \cdot 10^{-5})$~GeV/$c^2$. 
This demonstrates that the geometry of the DIRAC setup is well described. 

The width of the $\Lambda$ mass distribution allows testing the momentum and angular resolution of the setup in the simulation. 
The processed events were collected in the Data 1, Data 2, and Data 3 samples,
during three different runs. 
Table~\ref{tab:lambda-w} 
shows a good agreement between the simulated and experimental $\Lambda$ 
widths in Data 2 and Data 3.
A further test consists of comparing the experimental $\Lambda$ and 
${\bar{\Lambda}}$ widths. 

\begin{table}[htbp]
\scriptsize
\caption{The $\Lambda$ width in GeV/$c^2$  for the experimental and MC data 
and the ${\bar{\Lambda}}$ width for the experimental data.}
\label{tab:lambda-w}
\begin{center}
\begin{tabular}{|c|c|c|c|}
\hline 
\rule{0pt}{2.5ex} &   $\Lambda$ width (data)  &  $\Lambda$ width (MC)  & ${\bar{\Lambda}}$ width (data) \\
&      GeV/$c^2$     &     GeV/$c^2$   &     GeV/$c^2$ \\
\hline   
\rule{0pt}{2.5ex}  Data 2 &    $4.42 \!\cdot\! 10^{-4} \!\pm\! 7.4 \!\cdot\! 10^{-6}$ &  $ 4.42 \!\cdot\! 10^{-4} \!\pm\! 4.4 \!\cdot\! 10^{-6}$  &  $4.5 \!\cdot\! 10^{-4} \!\pm\! 3 \!\cdot\! 10^{-5}$ \\
\hline   
\rule{0pt}{2.5ex}  Data 3 &    $4.41 \!\cdot\! 10^{-4} \!\pm\! 7.5 \!\cdot\! 10^{-6}$ &  $ 4.37 \!\cdot\! 10^{-4} \!\pm\! 4.5 \!\cdot\! 10^{-6}$  &  $4.3 \!\cdot\! 10^{-4} \!\pm\! 2 \!\cdot\! 10^{-5}$ \\
\hline
\end{tabular}
\end{center}
\end{table}

The average value of the correction that was introduced in the simulated
width is $1.00203\pm0.00191\cdot10^{-3}$. 
Therefore, nonsignificant corrections were introduced in the l.s. particle momenta.


\subsection{Event selection} 
\label{ssec:Ev_sel}

Equal-mass pairs contained in the selected event sample are classified into three categories: $\pi^+\pi^-$, $K^+K^-$, and $p\bar{p}$ pairs.   

The classification is based on the TOF measurement \cite{note2001} 
for the distance between the SFD X-plane and the VH of about $11 m$.
For pairs with a total momenta range from $3.8$ to 8~GeV/$c$, 
additional information from the Heavy Gas Cherenkov (ChF) counters 
(Section~\ref{sec:setup}) is used to better separate $\pi^+\pi^-$ from 
$K^+K^-$ and $p\bar{p}$ pairs. The ChF counters detect pions in this region with (95--97)\% efficiency \cite{note1305}, whereas kaons and protons (antiprotons) do not generate any signal.

\section{Description of $\pi^+ \pi^-$ pair production and the simulation
procedure}
\label{sec:pair-prod}

The experimental distributions of the Coulomb and non-{\sl Coulomb pairs} in the relative momentum $Q$ and its components were compared with the corresponding simulated distributions.

The simulated Coulomb $\pi^+\pi^-$ spectra in the pair c.m.s. were 
calculated using the relation
\vspace{-1mm}
\begin{equation}
\label{dN-dQ}
\frac{dN}{dQ_i}=\left|M_\mathrm{prod}\right|^2 F(Q_i)\,A_C(Q_i)\,D(Q_i)
\end{equation}

where $Q_i$ is $Q, Q_L$ or $Q_t$, $M_\mathrm{prod}$ is the
production matrix element without $Q$ dependence in the
investigated $Q$ interval, $F(Q_i)$ is the phase space and
$A_C(Q)$ is the Coulomb correlation function
\vspace{-1mm}
\begin{equation}
\label{C-Corel-func}
A_C(Q)=\frac{2\pi m_\pi \alpha/Q}{1-exp(-2\pi m_\pi \alpha/Q)}
\end{equation}

with allowance for the Coulomb final state interaction (FSI). 

For small and large $Q$ the respective Coulomb correlation 
function values are 
\vspace{-1mm}
\begin{equation}
\label{C-Corel-Q-small-large}
A_C=2\pi m_\pi \alpha/Q \quad \mbox{and} \quad A_C=1
\end{equation}

The function $A_C(Q)$ in formula (\ref{C-Corel-func}) describes the Coulomb final state interaction of nonrelativistic particles. The function describing the same interaction of relativistic particles was evaluated in 
\cite{Arbuz94}.

The function $D(Q_i)$ in Eq. (\ref{dN-dQ}) takes into account small
  corrections caused by strong two-pion interaction in the final state
  and nonpoint-like pair production \cite{LEDN08}.
  The $D$-function dependence on $Q$ was calculated in \cite{LEDN08} using the space-time distribution of the pion production points based on
  the UrQMD transport code simulation \cite{BASS98},
  taking into account particle re-scatterings and resonance
  decays, including short-lived, intermediate ($\omega$), and long-lived
  ($\eta'$) resonances. Within possible uncertainties of the $\omega$
  and $\eta'$ fractions, it was shown that in the analyzed $Q$ interval
  up to 20 MeV/c, the correction $D$ function could be approximated as
  $D(Q)= c + b Q$ with $c= 1.01 - 1.06$ and $|b|<0.5/$GeV/$c$.
  The slope parameter $b$ strongly depends on the $\omega$
  and $\eta'$ fractions, changing from -0.5/GeV when they are ignored
  to +0.5/GeV when they are taken into account. 
  Detailed analysis and evaluation of various resonance contributions
  to the production of the $\pi^+ \pi^-$ at small $Q$ was done in 
\cite{CHLI09}.
  Since the estimated variation of the correction function $D(Q)$
  in the interval 0--20 MeV/$c$ is less than 1\%, one may describe
  the $Q$ distribution of the $\pi^+\pi^-$ {\sl Coulomb pairs} with this
  precision taking into account only the point-like Coulomb interaction.
  One can further improve this precision by taking into account the
  strong two-pion interaction and finite space-time separation of pion
  production points. In the present work, the correction function $D(Q)$
  from \cite{LEDN08} was used to describe the $Q_L$ and $Q_t$ distributions.

The same simulation was done for the non-Coulomb $\pi^+\pi^-$ pairs using formula (\ref{dN-dQ}) without the correlation function $A_C(Q)$.
The $Q_L$ distributions of non-Coulomb and {\sl accidental pairs} are the same.
Therefore, in all  ana\-lyses presented below the numbers of 
non-{\sl Coulomb pairs} include the contribution of {\sl accidental pairs}.

To calculate the momenta of the pair particles in the laboratory system 
(l.s.), the l.s. pair momentum is added to the c.m.s. one considering the difference between the total momentum distributions of the Coulomb and non-{\sl Coulomb pairs} in the l.s. \cite{ZHAB07,ZHAB08}.
This allows calculating the momenta $\vec P^+$ and $\vec P^-$
of the $\pi^+$  and $\pi^-$ in the l.s. and their total momentum
$\vec P=\vec P^++\vec P^-$.
By using the dedicated GEANT-DIRAC code, the simulated pairs are propagated through the setup with allowance for the multiple scattering and the response of the detectors in front of the magnet - the Scintillator Fiber Detector (SFD) and the Ionization Detector (ID).

The distance $D$ between two particles in the l.s. decreases with  $Q_t$ 
and for small $D$ in this experiment, the coordinate scintillation fiber 
detector with some probability cannot distinguish a one-particle hit 
from a two-particle hit. In this case, the amplitude is measured in 
the ionization detector. If the amplitude is higher than some threshold,  
this event is considered a two-particle hit. The introduction of the threshold results in rejecting part of the pairs and decreasing their detection efficiency $\epsilon$. This decrease begins with $D$ reducing below 0.8 mm in the $x$ and $y$ projections; the corresponding pair opening angle projections are 0.28 mrad.
Behind the spectrometer magnet, only events with one or two tracks per arm are selected.

On the basis of the information from the detectors, the events were 
reconstructed by the ARIANE code and processed as experimental pairs. 
The simulated event distribution in $\vec P_\mathrm{lab}$
was tuned by requiring that the Coulomb and non-{\sl Coulomb pairs}
fit the experimental $\pi^+\pi^-$ pair spectrum in 
$\vec P_\mathrm{exp}=\vec P_\mathrm{exp}^+ +\vec P_\mathrm{exp}^-$
where $\vec P_\mathrm{exp}^+$ and $\vec P_\mathrm{exp}^-$
are the experimental l.s. momenta of $\pi^+$ and $\pi^-$. 
After this the $Q_L, Q_t$ and $Q$ distributions of the simulated events 
were calculated and compared with the experimental spectra.

\section{Analysis of experimental $Q_L$ distributions of $\pi^+ \pi^-$ 
pairs and measurement of the number of Coulomb and non-Coulomb pairs}
\label{sec:exp-pi+pi-}
\vspace{-2mm}

For the analysis \cite{BENE23}, events with the time difference between the VH arms less than 0.5 ns were selected. These pair distributions in $Q_L$ were separately fitted in three data samples by a combination of the simulated {\sl Coulomb} and {\sl non-Coulomb pair} distributions in nine
$Q_t$ intervals: 0\,--\,0.25 (1), 0.25\,--\,0.5, 0.5\,--\,0.75, 0.75\,--\,1, 0 --\,1 (2), 1\,--\,2, 2\,--\,3 (3), 3\,--\,4 and 4\,--\,5~MeV/$c$ (4). Four of them are marked by numbers in parentheses for further reference.

\begin{figure*}[hbtp]
\centering
\hspace{-5mm}
\begin{tabular}{cc}
\includegraphics[width=67mm]{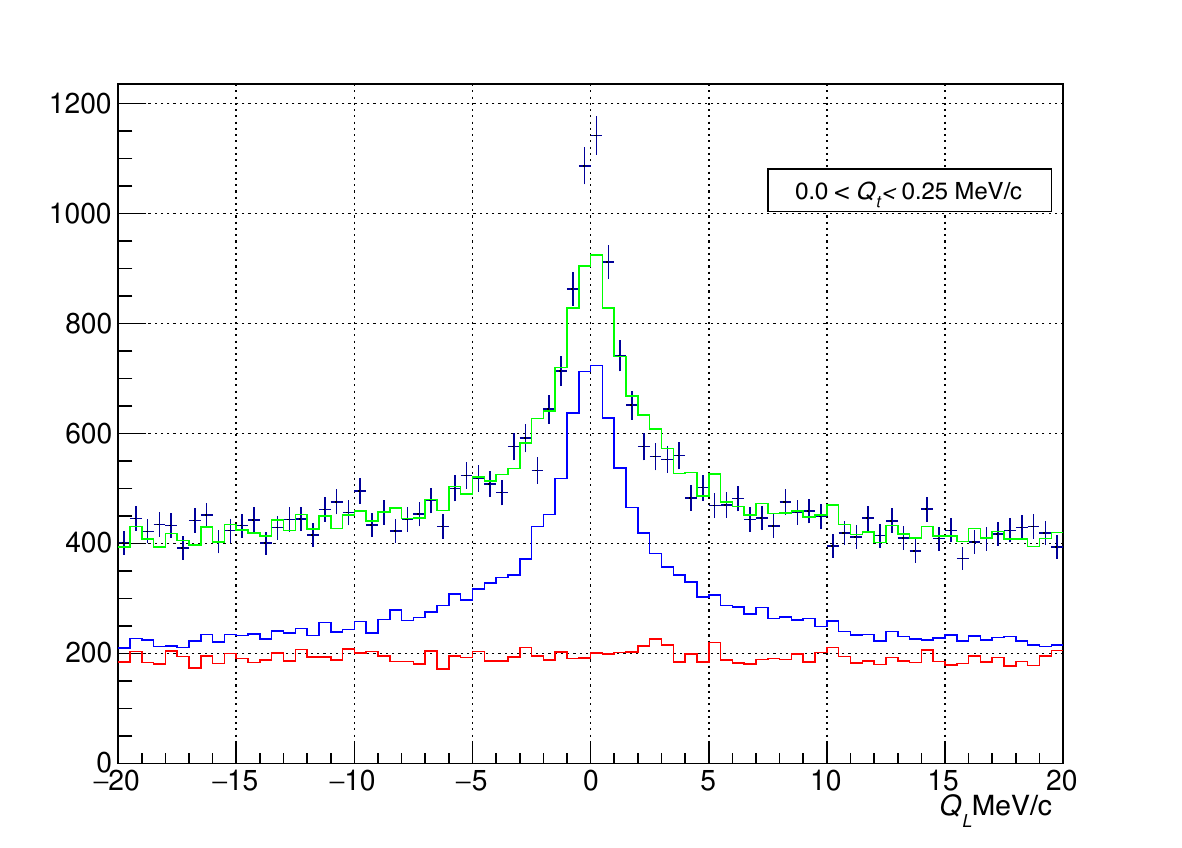}
&
\includegraphics[width=67mm]{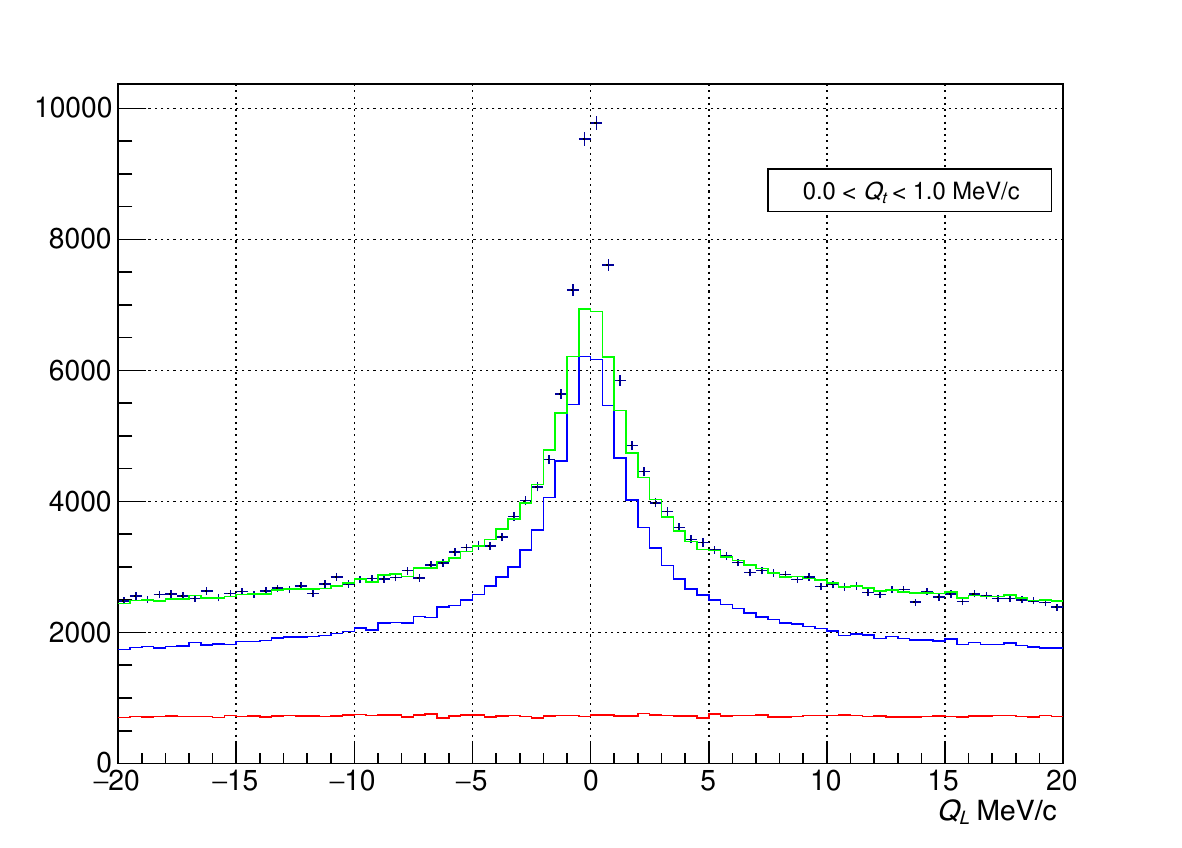}
\\
\includegraphics[width=67mm]{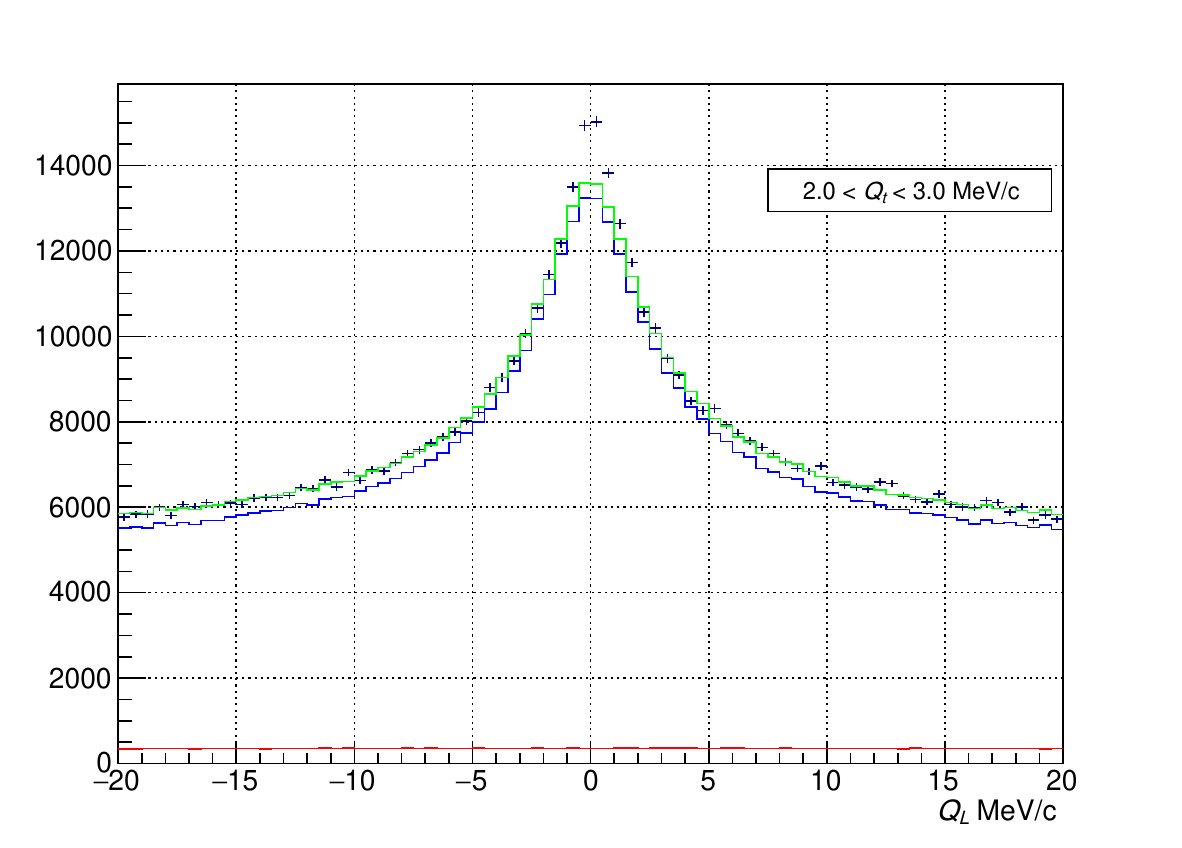}
&
\includegraphics[width=67mm]{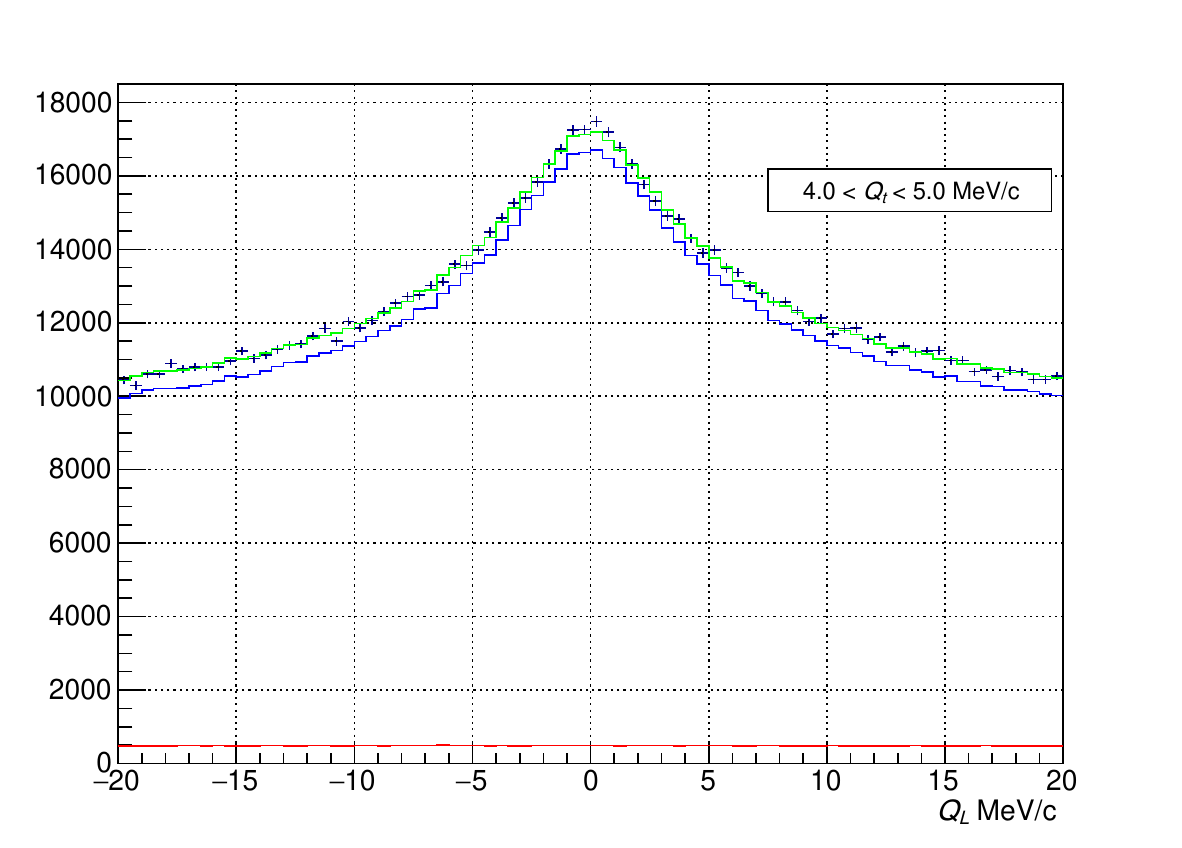}
\end{tabular}
\caption{
The experimental $Q_L$ distributions (black points with errors)
of the {\sl Coulomb}, {\sl non-Coulomb} and {\sl atomic $\pi^+\pi^-$ pairs} (sum of three data samples) for the $Q_t$ intervals 0.0\,--\,0.25~MeV/$c$, 0.0\,--\,1.0~MeV/$c$, 2.0\,--\,3.0~MeV/$c$ and 4.0\,--\,5.0~MeV/$c$. 
The green histogram is the corresponding combination of the {\sl Coulomb} and {\sl non-Coulomb pairs} simulated according to Eq. \ref{dN-dQ}. The fraction of the {\sl Coulomb pairs} and the normalization parameter were obtained by fitting in the total $Q_L$ interval except for the region -2~MeV/$c<Q_L<$ 2 MeV/$c$ populated by the {\sl atomic pairs}. One may see that the histograms well reproduce the increasing widths of the Coulomb peaks with increasing $Q_t$. The fitting histograms that describe the {\sl Coulomb} (blue) and {\sl non-Coulomb} (red) experimental pairs are presented as separate histograms. The excess pairs above the fitting histogram in the interval -2~MeV/$c<Q_L<$ 2~MeV/$c$ is due to the {\sl atomic pairs}.
}
\label{fig:QL-distrib}
\end{figure*}

\begin{figure*}[hbtp]
\hspace{-5mm}
\includegraphics[width=140mm]{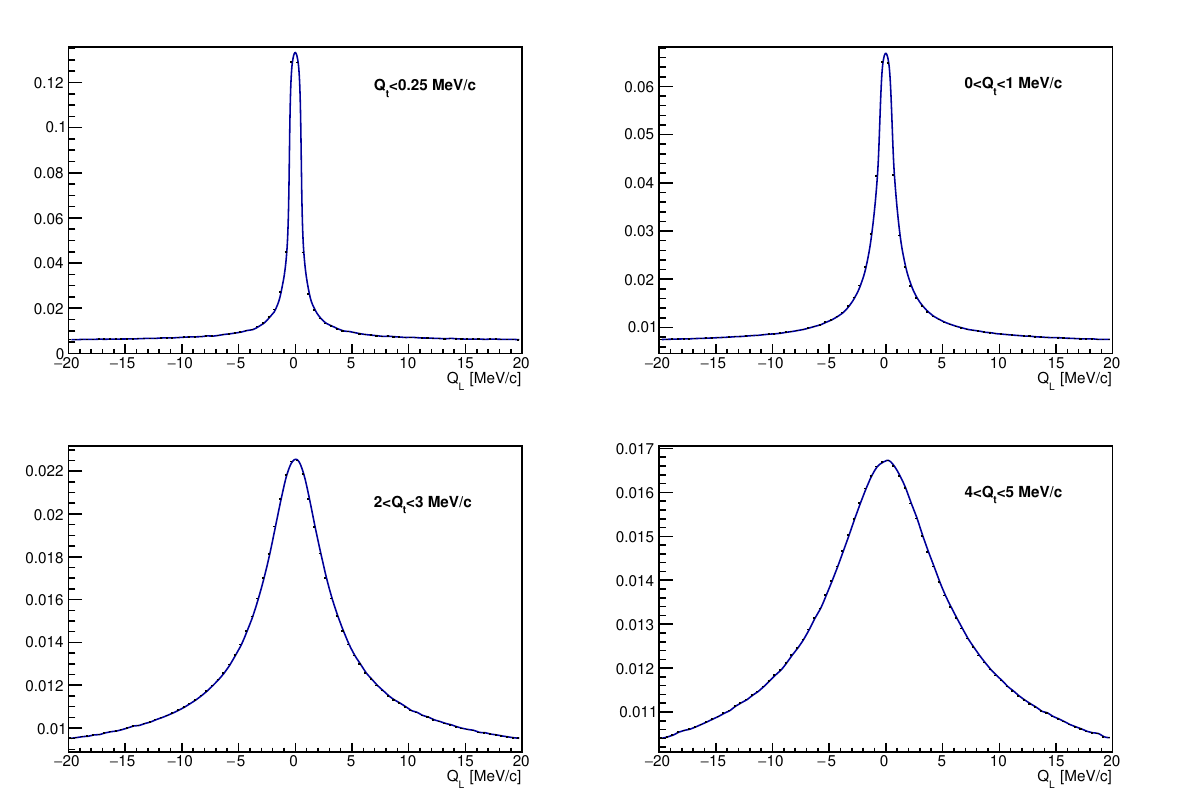}
\caption{
The simulated $Q_L$ distributions of $\pi^+\pi^-$ {\sl Coulomb pairs} at the production point for $Q_t$ intervals: 
~~0~--~0.25~MeV/$c$, ~~0~--~1~MeV/$c$, ~~2~--~3~MeV/$c$, 
~~4~--~5~MeV/$c$.
All distributions are normalized to unity.
It can be seen that the distribution width significantly increases with $Q_t$.
}
\label{fig:QL-sim-dis}
\end{figure*}

The ratio between these pair numbers in each interval was a free parameter. The fitting interval -20~MeV/$c < Q_L <$ 20~MeV/$c$ did not include the region -2~MeV/$c < Q_L <$ 2~MeV/$c$ which involves atomic pairs having different shapes of the $Q_L$ and $Q_t$ spectra. The number of simulated events for each data sample is an order of magnitude larger than the number of the corresponding experimental events. 

Figure \ref{fig:QL-distrib} shows the sum of three samples of 
experimental and fitting distributions in intervals (1) -- (4). 
Also, the fitting distributions of the {\sl Coulomb} and {\sl non-Coulomb pairs} are presented separately. The excess events in the interval 
-2~MeV/$c < Q_L <$ 2~MeV/$c$ are due to the atomic pairs. 
The experimental {\sl Coulomb pair} spectrum shows the peak
around $Q_L$ = 0. The full width at half maximum increases with
$Q_t$, and for $Q_t$ intervals 1, 2, 3, and 4 the width values are 
3.4~MeV/$c$, 4~MeV/$c$, 6.5~MeV/$c$ and 11~MeV/$c$ respectively.
They were obtained by measuring the histogram parameters.

The {\sl Coulomb pair} distributions in $Q_L$ for  $Q_t$
intervals (1) -- (4) at the pair production point were evaluated using
formula (\ref{dN-dQ}) and are presented in Fig. \ref{fig:QL-sim-dis}. 

The full width at half maximum for the four $Q_t$ intervals is 1.0~MeV/$c$, 1.2~MeV/$c$, 6.4~MeV/$c$, and 10.6~MeV/$c$ respectively. 
The same values for the experimental distributions in the $Q_t$ interval 0 -- 1~MeV/$c$ are significantly larger. In the DIRAC experiment, the main contribution to the width increase comes from the multiple scattering in the target. The multiple scattering in the detectors and the accuracy of the particle coordinate measurements are less important. 

Table \ref{tab-Ql-D1-D2-D3} shows the fitting procedure $\chi^2/ndf$ values for three data samples and five $Q_t$ intervals 

\begin{table}[h]
\vspace{-2mm}
\caption{}
\label{tab-Ql-D1-D2-D3}
\begin{tabular}{|c|c|c|c|c|c|}
\hline
\begin{minipage}{12mm}
\begin{center} $\Delta Q_t$ \\ (MeV/$c$)\end{center} 
\end{minipage} 
& 0 -- 1.0 & 1.0 -- 2.0 & 2.0 -- 3.0 & 3.0 -- 4.0 & 4.0 -- 5.0 \\ \hline
\multicolumn{6}{|c|}{$\chi^2/ndf$ values} \\ \hline 
Data 1 & 1.35 & 1.09 & 1.12 & 1.14 & 1.41 \\ \hline 
Data 2 & 1.45 & 0.90 & 1.29 & 0.90 & 1.19 \\ \hline 
Data 3 & 1.20 & 1.46 & 1.09 & 1.46 & 0.91 \\ \hline
\end{tabular}
\end{table}

\noindent
The $\chi^2/ndf$ values for four $Q_t$ intervals from 0\,--\,0.25 MeV/$c$ 
to 0.75\,--\,1 MeV/$c$ are presented in Table \ref{tab-QT-D1-D2-D}.
At the given $ndf\!=\!78$, the $\chi^2/{\rm ndf}$ probability distribution is close to the normal one with a mean of 1 and a standard deviation of 0.16. One may see from Tables \ref{tab-Ql-D1-D2-D3} and \ref{tab-QT-D1-D2-D} that the simulated $Q_L$ distributions fit the experimental ones quite well despite their strong widening with increasing $Q_t$. 
\vspace{-3mm}

\begin{table}[h]
\vspace{-3mm}
\small
\caption{}
\label{tab-QT-D1-D2-D}
\begin{tabular}{|c|c|c|c|c|}
\hline
\begin{minipage}{12mm}
\begin{center} $\Delta Q_t$ \\ (MeV/$c$)\end{center} 
\end{minipage} 
& 0 -- 0.25 & 0.25 -- 0.5 & 0.5 -- 0.75 & 0.75 -- 1.0 \\ \hline
\multicolumn{5}{|c|}{$\chi^2/ndf$ values} \\ \hline
Data 1 & 1.12 & 1.17 & 1.12 & 0.94 \\ \hline
Data 2 & 1.30 & 1.10 & 1.06 & 1.10 \\ \hline
Data 3 & 1.78 & 1.25 & 1.00 & 1.29 \\ \hline
\end{tabular}
\vspace{-3mm}
\end{table}

In each $Q_t$ interval that has a large $\chi^2/{\rm ndf}$ value for one data sample there are always two $\chi^2={\rm ndf}$ values for the other data samples less than 1.32 and only one value 1.35. 
A slight excess over unity of the averages of the values in Tables 
\ref{tab-Ql-D1-D2-D3} and \ref{tab-QT-D1-D2-D} may be due to uncertainties in the detector resolution and efficiency, as well as in the correction factor $D(Q)$ in Eq. (\ref{dN-dQ}).

\vspace{-3mm}
\begin{table}[h]
\small
\caption{}
\begin{tabular}{|c|c|c|c|c|c|}
\hline
\begin{minipage}{12mm}\begin{center} $\Delta Q_t$ \\ (MeV/$c$)\end{center} 
\end{minipage} 
& 0 -- 1.0 & 1.0 -- 2.0 & 2.0 -- 3.0 & 3.0 -- 4.0 & 4.0 -- 5.0 \\ \hline
$NC_\mathrm{exp}(\Delta Q_t)$ & 75200 & 137900 & 215900 & 298000 & 368600 
\\ \hline 
relative error & 3.6\% & 2.2\% & 1.9\% & 1.7\% & 1.8\% \\ \hline
$NC_\mathrm{calc}(\Delta Q_t)$ &  75720 & 140030 & 217330 & 294760 & 367190 \\ \hline
\end{tabular}
\label{table-3}
\vspace{-2mm}
\end{table}

Using fitted fractions of {\sl Coulomb pairs}, one may calculate their numbers $NC_{exp}(\Delta Q_t)$ - the experimental number of {\sl Coulomb pairs} and the corresponding relative errors in the intervals $\Delta Q_t$. Tables \ref{table-3} and \ref{table-4} show $NC_{exp}(\Delta Q_t)$ values for the Data 3 sample ($NC_{calc}(\Delta Q_t)$ will be defined in section \ref{sec:exp-distrib}).
It is seen that the relative precision of the number of {\sl Coulomb pairs} decreases with decreasing $Q_t$ because the background level becomes higher and the number of {\sl Coulomb pairs} becomes smaller. The relative errors in the Data 2 sample are the same as in the Data 3 sample. 

\begin{table}[h]
\vspace{-2mm}
\begin{center}
\caption{}
\begin{tabular}{|c|c|c|c|c|}
\hline
\begin{minipage}{12mm}\begin{center} $\Delta Q_t$ \\ (MeV/$c$)\end{center} 
\end{minipage} 
& 0 -- 0.25 & 0.25 -- 0.5 & 0.5 -- 0.75 & 0.75 -- 1.0 \\ \hline
$NC_\mathrm{exp}(\Delta Q_t$) & 9460 & 21760 & 21130 & 25750 \\ \hline
relative-error & 10.5\% & 6.3\% & 6.2\% & 5\% \\ \hline
$NC_\mathrm{calc}(\Delta Q_t)$ &  7890 & 19480 & 23190 & 25160 \\ \hline 
\end{tabular}
\label{table-4}
\end{center}
\vspace{-1mm}
\end{table}

The relative errors in the number of {\sl Coulomb pairs} depend on the description of the experimental conditions, statistical errors and simulated distribution precision.
In section IIB, reconstruction of the experimental $\Lambda$ and $\overline\Lambda$ masses and widths and their comparison with the same simulated parameters were described. It was shown that the average correction to the simulated event widths is at the level of about 0.2\%. It means that the setup geometry, momentum resolution and single-particle detection efficiency were defined well and the relative errors of the 
number of {\sl Coulomb pairs} give the minimum accuracy of the theoretical approach. 
A conclusion that can be drawn from the Data 2 and Data 3 analyses is that the theoreti\-cal approach using formula (\ref{dN-dQ}) allows one to describe the experimental distributions in $Q_L$ and to obtain the number of {\sl Coulomb pairs} with the precision better than 2\% in $Q_t$ intervals 2\,--\,3, 3\,--\,4 and 4\,--\,5~MeV/$c$.

It allows one to use Coulomb pairs to study and correct the quality of the simulation events. The standard procedure to check and correct the simulation event quality is to compare the experimental and simulated particle mass distributions. The Coulomb pair analysis provides an  additional possibility of checking the simulation accuracy. It has a special property which will be described in the next section.
\vspace{-2mm}

\section{Analysis of experimental $Q_t$ distributions of Coulomb pairs} 
\label{sec:exp-distrib}

In section \ref{sec:exp-pi+pi-} it was shown that the simulated distributions based on relation (\ref{dN-dQ}) described well $Q_L$ distributions of {\sl Coulomb pairs} for nine fixed  $Q_t$ intervals. In this section it will be shown \cite{BENE23} that formula (\ref{dN-dQ}) also describes the $Q_t$ distribution of the experimental {\sl Coulomb pairs} with $Q_L$ belonging to the interval -20~MeV/$c < Q_L <$ 20~MeV/$c$. If $Q_t$ decreases, the number of pairs with a small distance $D$ between the tracks increases in the corresponding $Q_t$ intervals. For these pairs, 
the detection efficiency $\epsilon$ has a strong dependence on $D$ 
(see section \ref{sec:pair-prod}), and errors in $\epsilon$ 
give rise to distortion of the number of simulated events and their greater 
difference from the number of experimental pairs. The analysis 
will also allow checking the accuracy of the $\epsilon$ dependence 
on $D$ used in the DIRAC simulation procedure.

The fitting procedure described in section \ref{sec:exp-pi+pi-} 
was applied to the experimental  $Q_L$  distribution of the pairs (fitting 
interval -20~MeV/$c<Q_L<$ 20~MeV/$c$,  excluding the region   
-2~MeV/$c<Q_L<$ 2~MeV/$c$) with the total $Q_t$ interval 
0\,--\,5~MeV/$c$ to obtain  $|M_\mathrm{prod}|^2$ for the simulated events and to evaluate the expected numbers of {\sl Coulomb pair} in different $Q_t$ intervals. 
The results of the analysis are presented in Table \ref{tab:Data-1-2-3}.

\begin{table}[htbp]
\caption{}
\label{tab:Data-1-2-3}
\scriptsize
\begin{center}
\vspace{-3mm}
\begin{tabular}{|p{26mm}|c|c|c|}
\hline 
\rule{0pt}{2.5ex} &   Data 1  &  Data 2  & Data 3 \\
\hline   
 \parbox[m]{5em}{Number~of Coulomb~pairs} & $710300$ &  $1108400$  &  $1095000$ 
\\
\hline   
\multicolumn{1}{|c|}{$\chi^2/ndf$} &    $1.4$ &  $1.01$  &  $1.4$ 
\\
\hline
\rule{0pt}{3.5ex}
\parbox[m]{5em}{Number~of~non-Coulomb~pairs}
&  35010  & 72340  &  $65260$
\\
\hline
\rule{0pt}{3.8ex}
 \parbox[m]{5em}{Ratio~${f=}$~{\sl Coulomb pairs}/total~pairs}
& $(95.3\pm 1.1)\%$  & $(93.9 \pm 0.9)\%$  &  $(94.4\pm 0.9)\%$
\\
\hline
\hspace{-3mm}
\begin{minipage}{28mm}
Ratio~$(1-f)=$non- Coulomb/total~pairs
\end{minipage}
& $(4.7\pm 1.1)\%$  & $(6.1\pm 0.9)\%$  & $(5.6\pm 0.9)\%$
\\
\hline
\end{tabular}
\end{center}
\vspace{-2mm}
\end{table}

It is seen in Table \ref{tab:Data-1-2-3} that the simulated 
distributions describe well all the experimental data in the $Q_t$ 
interval 0 -- 5 MeV/$c$. The fractions $f$ of the {\sl Coulomb pairs} in  
the three samples are in good agreement. 

The average fraction of {\sl Coulomb pairs} is $f=94.4\pm 0.5\%$, giving the average fraction of non-{\sl Coulomb pairs} $1-f=5.6\pm 0.5\%$.   
The analysis of the time spectra of {\sl prompt} and {\sl accidental pairs} gives for the relative contributions of {\sl accidental pairs} to the interval of 
prompt pairs $\pm 0.5~$ns, the values ($6.0 \pm 0.3$)\% (Data 2) and 
($6.2 \pm 0.6$)\% (Data 3). These numbers show that {\sl accidental pairs} make the main contribution to the {\sl non-Coulomb pairs}, leaving only a percent level window for the contribution of long-lived sources. Since the effect of $\eta'$ is taken into account in the factor $D(Q)$, this contribution is dominated by $\eta$-mesons, and their contribution is estimated to be less than a few percent \cite{CHLI09}.

\begin{figure}[hbtp]
\begin{center}
\includegraphics[width=90mm]{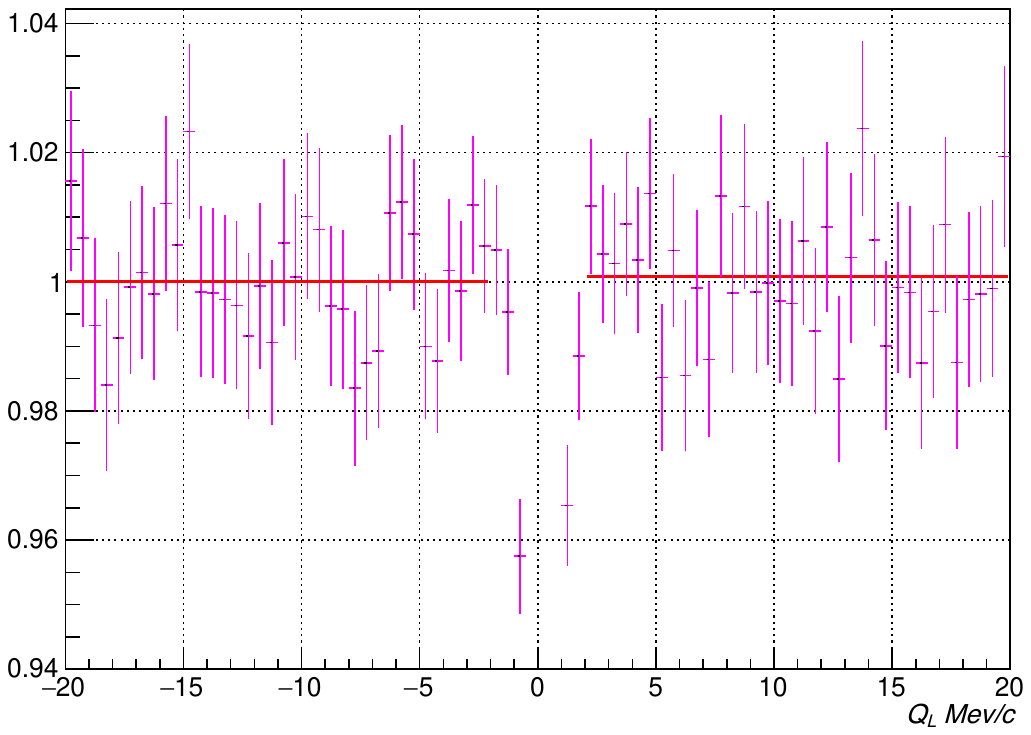}
\caption{
The simulated $Q_L$ distribution of the {\sl Coulomb} and {\sl non-Coulomb pairs} 
in each data sample was divided by the same experimental spectrum. 
The ratios for the three data samples  as a function of $Q_L$ were averaged and presented in this Figure. In the intervals of the positive and negative $Q_L$ (excluding  region $\pm$ 2MeV/$c$), the points were fitted independently by a 
constant. It is seen that in the left and right intervals, the average 
ratios are unity, demonstrating that 
formula (\ref{dN-dQ}) describes the $Q_L$ distribution well.
}
\label{fig:Data_MC}
\end{center}
\vspace{-4mm}
\end{figure}

The simulated distributions were obtained with the formula (\ref{dN-dQ}).
To further check the precision of equation (\ref{dN-dQ}), the simulated $Q_L$ distribution of the {\sl Coulomb} and {\sl non-Coulomb pairs} (in each data sample) was divided by the same experimental distribution.

The ratios for the three data samples as functions of $Q_L$ are averaged 
and presented in Fig. \ref{fig:Data_MC},
where for all the $Q_L$ values in the fitting intervals 
(excluding region $\pm$ 2~MeV/$c$) the ratios are about unity. 
The left and right sides of the ratios were fitted 
independently by a constant with a good $\chi^2$. 
The average ratio values for the negative and positive $Q_L$ are
$1.0000\pm 0.0021$ and $1.0009\pm 0.0021$ respectively \cite{BENE23}. 
One may conclude that formula (\ref{dN-dQ}) describes the $Q_L$ distribution of the experimental events with a precision better than 0.5\% for $Q_t$ in the interval 0 -- 5 MeV/$c$.

\begin{figure}[hbtp]
\hspace{-5mm}
\includegraphics[width=75mm]{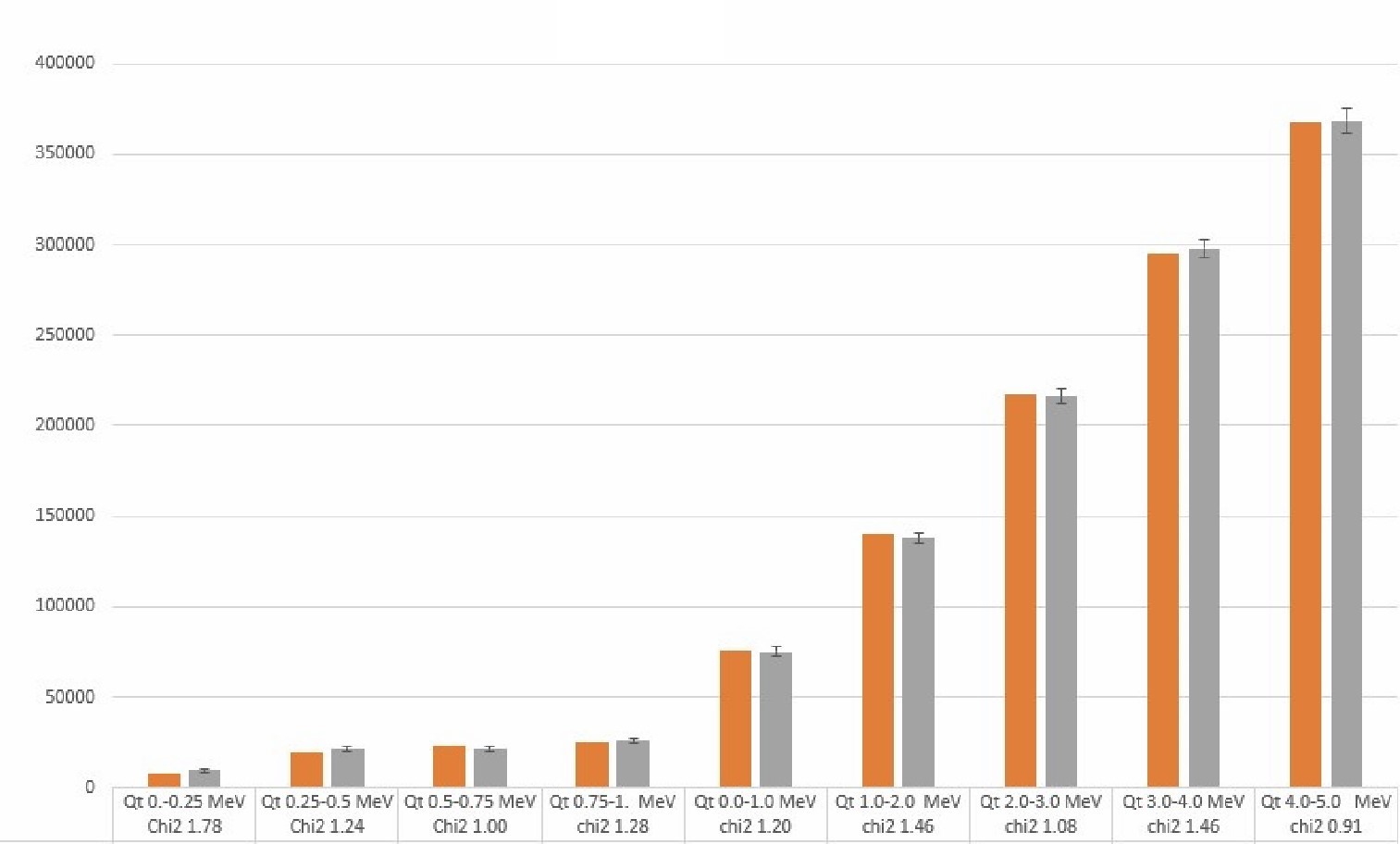}
\caption{The experimental numbers of {\sl Coulomb pairs} $NC_\mathrm{exp}(\Delta Q_t)$ for different $Q_t$ intervals (gray). The calculated numbers of {\sl Coulomb pairs}  $NC_\mathrm{calc}(\Delta Q_t)$ for the same $Q_t$ intervals (brown).
}
\label{fig:2010}
\vspace{-2mm}
\end{figure}


The $|M_\mathrm{prod}|^2$ evaluation allows calculating $NC_\mathrm{calc}(\Delta Q_t)$ the expected number of the simulated {\sl Coulomb pairs} in each of the $\Delta Q_t$ intervals analyzed in section \ref{sec:exp-pi+pi-}. 
The obtained $NC_\mathrm{calc}(\Delta Q_t)$ were compared with the 
$NC_\mathrm{exp}(\Delta Q_t)$ evaluated in section \ref{sec:exp-pi+pi-} 
by the fitting procedure in the same $\Delta Q_t$ intervals. 

The results of the Data 3 analysis for nine $\Delta Q_t$ intervals are presented in Fig.\ref{fig:2010} and in Tables \ref{table-3} and \ref{table-4}.
It is seen that the differences between these numbers in all nine $Q_t$ intervals are less than two standard deviations. 
The same good agreement is for the Data 1 and Data 2 samples. 

In section \ref{sec:exp-pi+pi-} it was shown that the fitting procedure using formula (\ref{dN-dQ}) described the experimental distributions in $Q_L$ 
with a precision better than 2\% in the $Q_t$ intervals 2\,--\,3, 3\,--\,4, and 4\,--\,5 MeV/$c$. 
The agreement between $NC_{calc}(\Delta Q_t)$  and $NC_{exp}(\Delta Q_t)$ demonstrates that formula (\ref{dN-dQ}) describes the experimental 
data in the same $Q_t$ intervals with a precision better than 2\% also.

Fig.\ref{fig:2010} shows that the main contribution to $|M_\mathrm{prod}|^2$  comes from the pairs with large $Q_t$. The distance $D$ of these pairs is large and the pair detection efficiency is well defined. The agreement between $NC_\mathrm{exp}(\Delta Q_t)$ and $NC_\mathrm{calc}(\Delta Q_t)$ for pairs with $Q_t$ less than 0.5~MeV/$c$ shows that the $\epsilon$ dependence on $D$ at a small distance between tracks was taken into account correctly.
\vspace{-2mm}

\section{The values of the angles between two tracks in the laboratory system}
\label{sec:5}
\vspace{-1mm}
 
The $Q_t$ values in the c.m.s. and the l.s. are the same. Therefore, the pairs with minimal $Q_t$, have the minimum opening angles $\theta$ and distance $D$ in the l.s.\!
The total momenta of the experimental pairs are found mainly in the interval 
2.4\,--\,8~GeV/$c$ with an average of about 4~GeV/$c$. 

In interval (4) the average $Q_t$ is 4.5~MeV/$c$. The angle between two particles in the l.s. at this $Q_t$ and the average total pair momentum is 2 mrad. 
The average $Q_t$ in interval (1) is around 0.12~MeV/$c$, and the opening angle for the average total pair momentum is 0.06 mrad. 
The contribution of the pairs with a smaller $Q_t$ and a larger total momentum
allows checking the detection efficiency for the pairs with the opening 
angles down to about 0.02 mrad.
The distribution with large $Q_t$ allows checking and correcting 
the simulation procedure for the pairs with large opening 
angles in the l.s. The detection efficiency for these pairs is the 
product of the well-known single-particle detection efficiencies. 
After the tuning of the simulation procedure and the evaluation 
of the $|M_\mathrm{prod}|^2$ value using these pair distributions,   
the expected numbers of the simulated {\sl Coulomb pairs} in the intervals with small $Q_t$ can be calculated with the theoretical accuracy better than 2\%. 
The comparison of the numbers of the simulated pairs $NC_\mathrm{calc}$ and the numbers of the experimental pairs $NC_\mathrm{exp}$  
allows checking and correcting the detection efficiency for 
the pairs with a small distance $D$ \cite{BENE23}.
This possibility is the particular pro\-perty of the method using Coulomb pairs.
\vspace{-2mm}

\section{Conclusion}
\vspace{-1mm}

In this work, the Coulomb effects in the $\pi^+\pi^-$ pairs were studied 
and their application to the data processing is justified. 
The $\pi^+\pi^-$ pairs were produced in p-Ni interactions with the proton 
momentum of 24~GeV/$c$. 
The experimental data samples Data 1, Data 2 and Data 3 were obtained in three different runs. 
The Coulomb effects (Coulomb correlations) were studied using 
the experimental pair distributions in $Q$, the relative momentum 
in the pair c.m.s., and its longitudinal ($Q_L$) and transverse ($Q_t$) 
projections on the pair direction in the l.s. 
The major part of the $\pi^+~\pi^-$ pairs was produced in decays of $\rho, \omega, \Delta$ and other short-lived resonances ({\sl Coulomb pairs}). In these pairs at 
small $Q$, the significant Coulomb interaction in the 
final state arises and increases the pair yield with decreasing $Q$. 
The minor part of the pairs contains one or both pions 
resulting from long-lived sources, such as $\eta, \eta'$ or from different 
events ({\sl non-Coulomb pairs}). 
In this case, the distance between particles would be much larger 
than the Bohr radius of the $\pi^+\pi^-$ atom, and the pion interaction 
in the final state is practically or completely absent. 

The experimental $\pi^+\pi^-$ pair distributions were analyzed in 
the intervals $0\!<\!Q_t\!<\!5$~MeV/$c$ and $-20$~MeV/$c <Q_L<$~20~MeV/$c$ using a 
combination of the corresponding simulated {\sl Coulomb} and {\sl non-Coulomb pair}  
distributions. The simulated spectra of the {\sl Coulomb} $\pi^+\pi^-${\sl pairs} 
in the c.m.s. were obtained 
according to (\ref{dN-dQ}) with taking into account the $\pi \pi$ Coulomb and strong interaction in the final state and the nonpoint-like pair production.

The {\sl non-Coulomb pairs} were simulated according to Eq.(\ref{dN-dQ}) without correlation, with $A_C(Q)\!=\!D(Q)\!=\!1$.

All experimental events were divided into nine $Q_t$ intervals:  
0\,--\,0.25, 0.25\,--\,0.5, 0.5\,--\,0.75, 0.75\,--\,1, 0\,--\,1, 1\,--\,2, 2\,--\,3, 3\,--\,4 and 4\,--\,5~MeV/$c$.  
In each interval, $Q_L$ spectra were obtained, which showed peaks 
around $Q_L = 0$ caused by the Coulomb final state interaction 
(Fig. \ref{fig:QL-distrib}). 

Each distribution was fitted (section \ref{sec:exp-pi+pi-}) by a 
combination of the simulated {\sl Coulomb} and {\sl non-Coulomb pairs} with two free parameters: 
the fraction of {\sl Coulomb pairs} and the normalization constant. 
The fitting interval did not include the region $-2$~MeV/$c<Q_L<$ 2~MeV/$c$ which contains atomic pairs that arose from the breakup of  $\pi^+\pi^-$ atoms in the target (Fig. \ref{fig:atom-production}) and had a different shape of the $Q_L$ and $Q_t$ spectra.  

Nine experimental distributions in all three data samples were 
described well. The full width at half maxi\-mum increases with $Q_t$ 
and is 3.4~MeV/$c$ ($0\!<\!Q_t\!<\!0.25$~MeV/$c$), 4~MeV/$c$ ($0\!<\!Q_t\!<\!1$~MeV/$c$), 
6.5~MeV/$c$ ($2\!<\!Q_t\!<\!3~MeV/$c$)$ and 11~MeV/$c$ ($4\!<\!Q_t\!<\!5$~MeV/$c$).
It appears that Eq. (\ref{dN-dQ}) describes the numbers of
{\sl Coulomb pairs} in $Q_t$ intervals 2\,--\,3, 3\,--\,4 and 4\,--\,5~MeV/$c$ 
with a precision better than 2\%.

A dedicated analysis of the precision of Eq.(\ref{dN-dQ}) was done in the
total $Q_t$ interval 0\,--\,5~MeV/$c$ using the fitting procedure described in 
section \ref{sec:exp-pi+pi-}. The fitted combi\-nation of the simulated 
$Q_L$ distributions in each data sample was divided by the corresponding 
experimental spectrum. The averaged ratios for the three data samples were fitted by a constant, separately for positive and negative $Q_L$, excluding the region $Q_L<2$ MeV/$c$ (Fig. \ref{fig:Data_MC}). The corresponding fitted ratios, $1.0000\pm 0.0021$ and $1.0009 \pm 0.0021$, and good quality of the fits demonstrate that the simulation procedure based on Eq. (\ref{dN-dQ}) describes the experimental $Q_L$ distributions with the precision better than $0.5\%$.

The evaluated $|M_\mathrm{prod}|^2$ in the $Q_t$ interval 0\,--\,5~MeV/$c$ allow the calculation of $NC_\mathrm{calc}(\Delta Q_t)$, the expected number of the simulated {\sl Coulomb pairs} in each of the nine $\Delta Q_t$ intervals. 
The obtained $NC_\mathrm{calc}(\Delta Q_t)$ values were compared with $NC_\mathrm{exp}(\Delta Q_t)$, the experimental numbers evaluated by the fitting procedure in the same $\Delta Q_t$ intervals (Fig. \ref{fig:2010}).

It is shown that in the three data samples and in all nine $Q_t$ intervals there is a good agreement between 
$NC_\mathrm{exp}(\Delta Q_t)$ and the number of the simulated events 
$NC_\mathrm{calc}(\Delta Q_t)$. 

It demonstrates, together with the good $\chi^2$, that formula 
(\ref{dN-dQ}) describes the experimental $Q_t$ and $Q_L$ 
distributions of {\sl Coulomb pairs} with a precision of better 
than 2\% and the dependence of the two-particle detection efficiency $\epsilon$ on the 
distance $D$ between the particles is taken into account correctly.

The pairs with the minimal $Q_t$ have the minimum opening angles $\theta$ and the minimum distance $D$ in the laboratory system. 
The total momenta of the experimental pairs are mainly in the 
interval 2.4~GeV/$c$ -- 8~GeV/$c$ 
with an average value of about 4~GeV/$c$. 

At $Q_t$\,=\,4.5~MeV/$c$ (interval 4.0\,--\,5.0~MeV/$c$) and the total 
momentum of 4.0~GeV/$c$, the angle $\theta$ is 2 mrad. 
At $Q_t$\,=\,0.12~MeV/$c$ (interval 0\,--\,0.25~MeV/$c$) the 
corresponding opening angle is 0.06 mrad. In this $Q_t$ interval, 
there is a significant number of the simulated events with 
smaller $Q_t$ and larger total momenta in the l.s. 
These pairs allow checking the detection efficiency for the pairs
with the opening angles down to 0.02 mrad.

Finally, we have shown that the selection of $\pi^+ \pi^-$ {\sl Coulomb pairs} in different $Q_t$ intervals less than 5 MeV/$c$ allows one to form $Q_L$ distributions of pairs with peaks around $Q_L=0$ and with different widths. 
These distri\-butions can be described with a theoretical precision better than 2\% in the $Q_t$ intervals 2\,--\,3, 3\,--\,4 and 4\,--\,5 MeV/$c$. In the same $Q_t$ intervals,
the number of {\sl Coulomb pairs}  
can be calculated with the 2\% accuracy. It is shown that for $Q_t$ in the interval 0\,--\,5 MeV/c formula (\ref{dN-dQ}) describes the $Q_L$ distribution of the experimental events with a precision better than 0.5\%.
The ordinary way to investigate the quality of the simulated events is based on comparing the reconstructed and simulated particle mass distributions. The properties of the Coulomb pairs allow one to use these pairs as a new physical tool to check and correct the simulated event quality. The particular property of the {\sl Coulomb pairs} is the possibility of checking and correcting the detection efficiency for the pairs with small opening angles.

\end{document}